\documentclass[letter]{aa} % for the letters 
\usepackage{aas_macros}
\usepackage{graphicx}
\usepackage{txfonts}
\usepackage{hyperref}
\graphicspath{{Images/}{../Images/}}
\usepackage{subfiles}
\usepackage{xcolor}
\usepackage[normalem]{ulem}

\def\Ystar{Y^*}
\def\Xstar{X^*}
\def\Ytilde{\tilde{Y}}
\def\kB{k_{\rm B}}

\begin{document}

    \title{Accounting for non-ideal mixing effects in the hydrogen-helium equation of state\thanks{Table B.1. is available via \url{https://doi.org/10.5281/zenodo.7346181}.}}

    \author{S. Howard \inst{1}
          \and T. Guillot\inst{1}
          }

    \institute{Université Côte d'Azur, Observatoire de la Côte d'Azur, CNRS, Laboratoire Lagrange, France\\
              \email{saburo.howard@oca.eu}
             }

    \date{Accepted February 2, 2023}
    
% \abstract{}{}{}{}{} 
% 5 {} token are mandatory
 
  \abstract
   {The equation of state for hydrogen and helium is fundamental for studying stars and giant planets. It has been shown that because of interactions at atomic and molecular levels, the behaviour of a mixture of hydrogen and helium cannot be accurately represented by considering these elements separately.
   }
   {This paper aims at providing a simple method to account for interactions between hydrogen and helium in interior and evolution models of giant planets.  
   }
   {Using on the one hand ab initio simulations that involve a system of interacting hydrogen and helium particles and pure equations of state for hydrogen and helium on the other, we derived the contributions in density and entropy of the interactions between hydrogen and helium particles.}
   {We show that relative variations of up to 15\% in density and entropy arise when non-ideal mixing is accounted for. These non-ideal mixing effects must be considered in interior models of giant planets based on accurate gravity field measurements, particularly in the context of variations in the helium-to-hydrogen ratio. They also affect the mass-radius relation of exoplanets.  
   We provide a table that contains the volume and entropy of mixing as a function of pressure and temperature. This table is to be combined with pure hydrogen and pure helium equations of state to obtain an equation of state that self-consistently includes mixing effects for any hydrogen and helium mixing ratio and may be used to model the interior structure and evolution of giant planets to brown dwarfs.
   }
   {Non-linear mixing must be included in accurate calculations of the equations of state of hydrogen and helium. Uncertainties on the equation of state still exist, however. Ab initio calculations of the behaviour of the hydrogen-helium mixture in the megabar regime for various compositions should be performed in order to gain accuracy.}


   \maketitle
%
%-------------------------------------------------------------------
%-------------------------------------------------------------------
\section{Introduction}
%-------------------------------------------------------------------
%-------------------------------------------------------------------

The hydrogen and helium equation of state (hereafter H-He EOS) is crucial in many contexts, but is particularly challenging to model in a high-pressure, low-temperature regime where interactions between molecules, atoms and ions are substantial, near a megabar and at temperatures lower than $10^4$\,K \citep[e.g.][]{2018oeps.book..175H}. These conditions concern giant planets \citep{2005AREPS..33..493G,2010exop.book..397F,2016JGRE..121.1552M} and brown dwarfs in particular \citep{2014ASSL..401..141B}.
With pressures reaching up to terabar (for the most massive and cold brown dwarfs), these objects span a range of thermodynamical conditions that includes low-pressure regimes for which experimental data are available and high-pressure regimes that are reasonably well understood \citep{2020NatRP...2..562H}, but, with temperatures in the megabar regime ranging from about 5000\,K to 20,000\,K \citep{2015trge.book..529G}, they cross this difficult and uncertain regime in parameter space. Understanding their interior structure and evolution requires an accurate EOS relating physical parameters such as pressure, temperature, density and entropy. 

Except for the specific case of the hydrogen-helium phase separation \citep[e.g.][]{2021Natur.593..517B}, most of the effort has been invested so far in the analysis of pure systems of either hydrogen or helium. This has been realised by means of laboratory experiments with anvil cells and laser-driven shocks, and theoretically, with approaches such as density functional theory (DFT), path integral Monte Carlo (PIMC) and quantum Monte Carlo (QMC) based on first-principles simulations (see \citet{2020NatRP...2..562H} for more details). The pure EOSs are usually coupled with the linear mixing rule (also known as the additive volume rule), whose principle is to consider that for extensive variables such as volume and entropy, the value of the mixture is equal to the sum of the pure species weighted by their respective abundances \citep{1995ApJS...99..713S}. 
However, \citet{PhysRevB.75.024206} pointed out that in the regime in which hydrogen transitions from a molecular to an atomic fluid, the presence of helium shortens and strengthens the molecular hydrogen bonds, leading to non-linear effects and potentially non-negligible deviations from the linear mixing hypothesis. This implies that in addition to pure EOSs, the behaviour of mixtures should be calculated for all values of the abundances considered. Some DFT simulations have been conducted to study the conditions for a phase separation of hydrogen and helium, but only for a limited range of parameters \citep{Morales2013,SR2018_PRL,SR2018_JPl} (hereafter called SR18). An extensive DFT calculation for a H-He mixture has been carried out by \citet{2013ApJ...774..148M} to provide an EOS (the so-called MH13 EOS) that fully accounts for the interactions between hydrogen and helium particles for temperatures between 1000 and 80,000\,K and pressures between 0.1 and 300\,Mbar. However, it is available only for one specific composition, namely a mixture of 220 hydrogen atoms for 18 helium atoms, corresponding to a helium mass-mixing ratio $Y=0.245$. Extensions of this table to cover a wider range of pressures and temperatures have been made available by \citet{2016A&A...596A.114M} and \citet{2021ApJ...917....4C} (CD21, hereafter). However, while these works provide EOSs that in principle can be applied to any mixture of hydrogen and helium, the treatment of non-linear effects remains fixed and equal to that calculated by \citet{2013ApJ...774..148M} at $Y=0.245$.

Our work aims to obtain a H-He EOS that includes the hydrogen and helium interactions and remains valid for any hydrogen-to-helium ratio. This EOS should recover the CD21 EOS as well as both the pure end members CMS19-H and CMS19-He \citep{2019ApJ...872...51C}, when there is only hydrogen or only helium in the system, respectively. Section~\ref{section:derivation} describes how the table is built that contains the contribution of the interactions between hydrogen and helium. Sections~\ref{section:jupiter} and~\ref{section:exoplanets} are dedicated to applications to Solar System planets and exoplanets, respectively.

%-------------------------------------------------------------------
%-------------------------------------------------------------------
\section{Derivation of the non-ideal mixing effects}
  \label{section:derivation}
%-------------------------------------------------------------------
%-------------------------------------------------------------------

In order to account for the interactions between hydrogen and helium, we make use of the linear mixing rule, but add a term $\Delta V$ or $\Delta S$ that corresponds to an excess or deficit of volume or entropy, respectively,
\begin{equation}
    \frac{1}{\rho_{\rm H-He}} = \frac{X}{\rho_{\rm H}} + \frac{Y}{\rho_{\rm He}} + \Delta V(X,Y),
    \label{eq:avl_rho_dv}
\end{equation}
\vspace*{-0.5cm}
\begin{equation}
    S_{\rm H-He} = XS_{\rm H} + YS_{\rm He} + \Delta S(X,Y),
    \label{eq:avl_s_ds}
\end{equation}
where $X$ and $Y$ are the mass fractions of hydrogen and helium, $\rho_{\rm H}$, $\rho_{\rm He}$ and $\rho_{\rm H-He}$ are the densities of pure hydrogen, pure helium, and the hydrogen-helium mixture, respectively, and $S_{\rm H}$, $S_{\rm He}$ , and $S_{\rm H-He}$ are the specific entropies of pure hydrogen, pure helium, and the hydrogen-helium mixture, respectively. The volume of mixing $\Delta V$ and entropy of mixing $\Delta S$ depend on $X$ and $Y$, namely the composition of the mixture. All quantities are evaluated at a given pressure $P$ and temperature $T$. A non-zero volume of mixing results from interactions between hydrogen and helium. The entropy of mixing is the sum of a non-ideal part and an ideal part $\Delta S_{\rm ideal}$ , arising even without these interactions. 
Following \citet{2019ApJ...872...51C} \citep[see also][]{1995ApJS...99..713S}, we write
\begin{equation}
    \Delta S_{\rm ideal} = - \kB \frac{x_{\rm H}{\rm ln}(x_{\rm H})+x_{\rm He}{\rm ln}(x_{\rm He})}{\langle A \rangle m_{\rm H}},
    \label{eq:ds_analytic}
\end{equation}
where $\kB$ refers to the Boltzmann constant, $x_i$ is the number fraction of component i, $m_{\rm H}$ is the atomic mass unit, and $\langle A \rangle=\sum{x_i A_i}$, $A_i$ being the mass number. 

Given the  known EOSs for pure hydrogen, for pure helium, and for one mixture of hydrogen and helium with mass mixing ratios $X^*= 1-\Ystar$ and $\Ystar$, respectively, a natural choice is to assume as a first-order approximation, 
\begin{equation}
    \Delta V(X,Y) = XYV_{\rm mix}; \qquad \Delta S(X,Y) = XYS_{\rm mix}.
    \label{eq:XYVmix}
\end{equation}
$V_{\rm mix}$ and $S_{\rm mix}$ are then quantities that are independent of $X$ and $Y$ (but vary as a function of $P$ and $T$). Using Eqs.~\eqref{eq:avl_rho_dv} and \eqref{eq:avl_s_ds}, we can hence calculate $V_{\rm mix}$ and $S_{\rm mix}$, 
\begin{equation}
    V_{\rm mix} = \frac{1}{\tilde{X}\tilde{Y}}\left[
    \frac{1}{\rho_{\rm H-He^*}} -
    \frac{\Xstar}{\rho_{\rm H}} - \frac{\Ystar}{\rho_{\rm He}} \right],
    \label{eq:vmix}
\end{equation}
\vspace*{-0.5cm}
\begin{equation}
    S_{\rm mix} = \frac{1}{\tilde{X}\tilde{Y}}\left[
    S_{\rm H-He^*} - \Xstar S_{\rm H} - \Ystar S_{\rm He} - \Delta S_{\rm ideal}(\Ystar) + \Delta S_{\rm ideal}(\Ytilde) \right].
    \label{eq:smix}
\end{equation}
Here we have included a complication that arises when the table for the mixture is calculated for a helium mass-mixing ratio $\Ystar$, but the mixing terms have been evaluated for a different mixing ratio $\Ytilde$. In the case that we consider, the CD21 H-He table is given for a helium mass-mixing ratio $\Ystar=0.275,$ but includes the non-ideal mixing effects from \citet{2013ApJ...774..148M}, who evaluated the H-He interactions for a mixture with $\Ytilde=0.245$. Equation.~\eqref{eq:smix} also accounts for the fact that the CD21 EOS evaluates the ideal entropy of mixing at $\Ystar=0.275,$ but the non-ideal entropy of mixing at $\Ytilde=0.245$.

\begin{figure}
   \centering
   \includegraphics[width=\hsize]{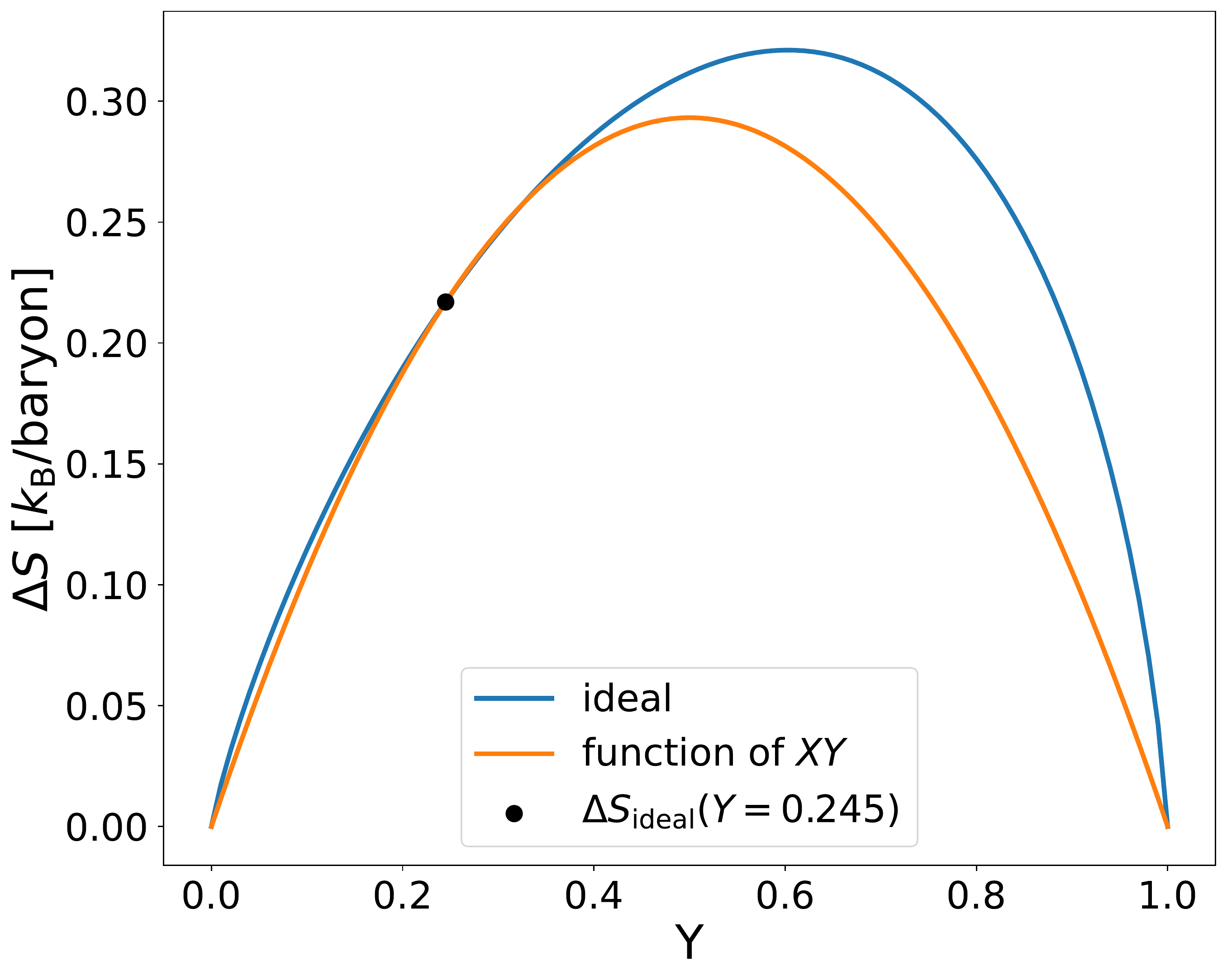}
      \caption{Comparison of the ideal entropy of mixing (blue curve and Eq.~\eqref{eq:smix}) with its first-order approximation, assumed proportional to $XY$, the product of the mass-mixing ratios of hydrogen and helium (orange curve), calculated as $\Delta S_{\rm ideal}(\Ytilde) \ XY/(\tilde{X}\tilde{Y})$, with $\Ytilde=0.245$.
              }
         \label{figure:smix_xy}
\end{figure}

We note that Eq.~\eqref{eq:ds_analytic} yields a value of $\Delta S_{\rm ideal}$ that is not strictly proportional to $XY$, in contrast to the assumption used in Eq.~\eqref{eq:XYVmix}. However, Fig.~\ref{figure:smix_xy} shows that the error made by approximating the ideal entropy of mixing as if it were proportional to $XY$ is small: it is always smaller than $0.1\,\kB$/baryon, and when we limit ourselves to compositions such that $Y<0.4$, the error made is smaller than $0.01\,\kB$/baryon. This thus supports our assumption that mixing terms should to first order be proportional to $XY$. 

\begin{figure}
   \centering
      {\includegraphics[width=\hsize]{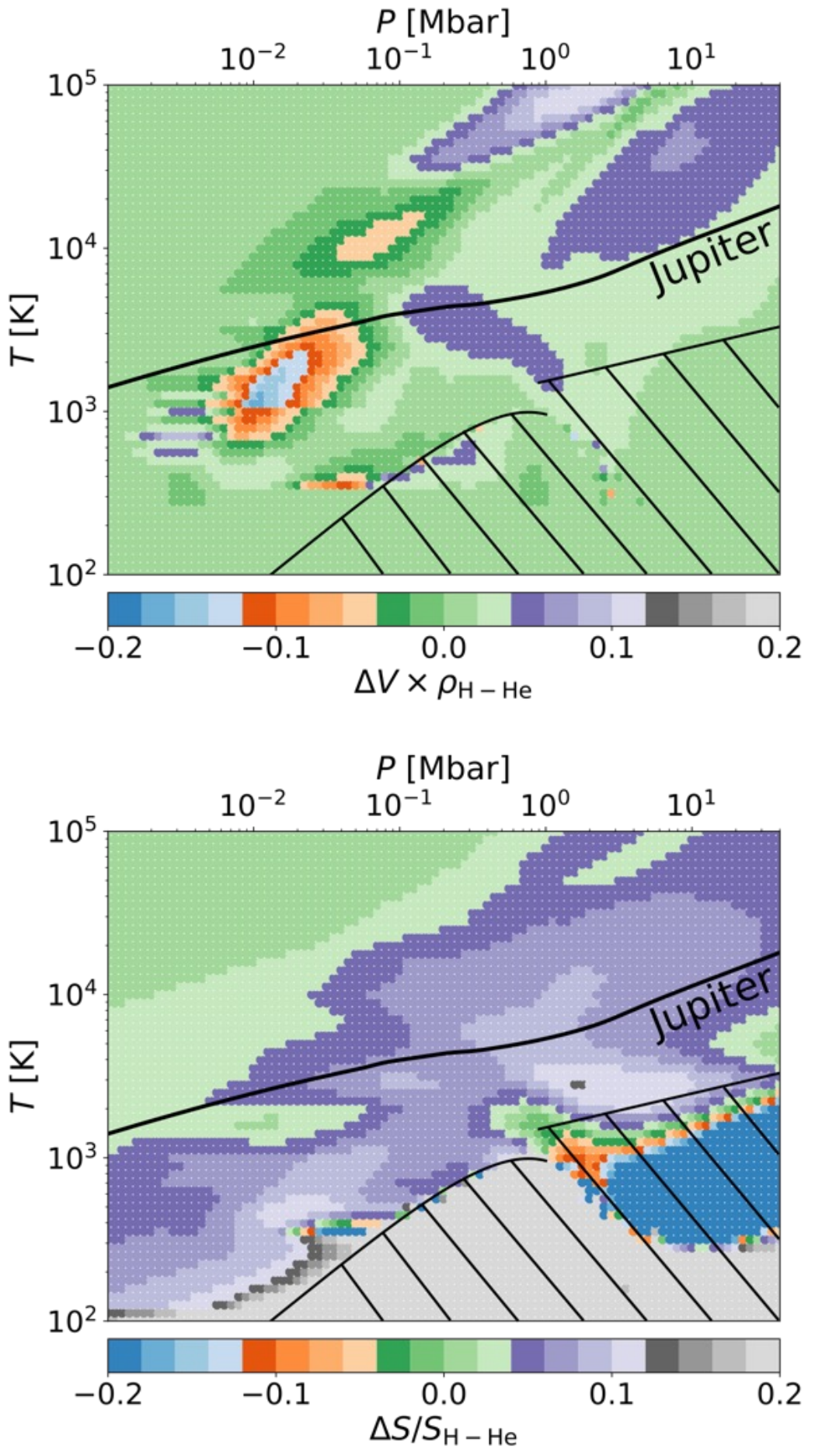}}
      \caption{Relative contributions of the mixing terms $\Delta V$ and $\Delta S$ to the total specific volume $1/\rho_{\rm H-He}$ and entropy $S_{\rm H-He}$, respectively, as a function of pressure and temperature. The solid black line corresponds to the Jupiter adiabat from \citet{2022A&A...662A..18M}. The values of $\Delta V\times \rho_{\rm H-He}$ and $\Delta S/S_{\rm H-He}$ are calculated for $\Xstar=0.725$ and $\Ystar=0.275$.
      The hashed areas correspond to the region in which the mixing volume and entropy are invalid \citep[see][for more details]{2019ApJ...872...51C}.
      }
         \label{figure:dv_ds}
\end{figure}

Using Eqs.~\eqref{eq:vmix} and \eqref{eq:smix}, the H-He table from \citet{2021ApJ...917....4C}, and the pure H and He tables from \citet{2019ApJ...872...51C}, which all use the same $P-T$ grid (pressures from $10^{-11}$ to $10^{11}$~Mbar and temperatures from $10^{2}$ to $10^{8}$~K), we can calculate a new table for $V_{\rm mix}$ and $S_{\rm mix}$ for the same set of pressures and temperatures. We stress that this table is valid in the same domain as the EOSs from \citet{2019ApJ...872...51C} and \citet{2021ApJ...917....4C}, whose limitations concern the regions in which molecular hydrogen and ionised hydrogen become solid as well as the region in which ion quantum effects become important, namely at low T and high P (see Fig.1 and Fig.16 of \citet{2019ApJ...872...51C} for the precise locations in the phase diagrams). The table was cleaned as described in Appendix~\ref{section:app2} to avoid spurious numerical effects and is available for download. Combined with the CMS19 tables, it can be used to calculate the internal structure of giant planets and brown dwarfs of variable compositions and including non-ideal mixing effects. The EOS for any mixture may then be evaluated from the EOSs for the pure elements, Eqs.~\eqref{eq:avl_rho_dv} and \eqref{eq:avl_s_ds}, $\Delta V=XY V_{\rm mix}$ and $\Delta S=XY S_{\rm mix}$. 

Figure~\ref{figure:dv_ds} shows the relative amplitude of the mixing quantities $\Delta V$ and $\Delta S$ in a mixture with $Y=0.245$. A typical Jupiter adiabat passes a region between 0.01~Mbar and 0.04~Mbar, where the non-ideal mixing effects are strongest. In this region, the density of the hydrogen and helium mixture inside Jupiter is denser by almost 10\% compared to the density obtained without accounting for the H-He interactions. Around 0.06~Mbar, the sign of $\Delta V$ changes and becomes positive (the mixture becomes less dense). Then, $\Delta V$ continues to increase to approximately 0.2~Mbar and starts to decrease afterwards. 

The variations in $\Delta V$ obtained in the $0.1-1\,$Mbar pressure range for Jupiter conditions are in line with those obtained by \citet{PhysRevB.75.024206}. However, we point out that we obtain unanticipated negative contributions (i.e. non-linear mixing terms yielding higher densities) at lower pressures. We note that this low-pressure region, around $0.01\,$Mbar, is a sensitive area in which EOS tables are combined, possibly yielding interpolation errors.

The bottom panel of Fig.~\ref{figure:dv_ds} shows that the presence of helium yields a mixing term that is small in the classical regime at low pressures and high-enough temperatures, but reaches up to 10\% away from it. On the Jupiter adiabat, this increase in $\Delta S$ occurs at pressures close to 0.1\,Mbar, peaks in the megabar region, and slowly decreases at higher pressures. For temperatures that are about five times lower than on the Jupiter adiabat, we enter the invalidity regime of the EOSs. Results obtained in this regime should not be trusted.

\begin{figure}
   \centering
      {\includegraphics[width=\hsize]{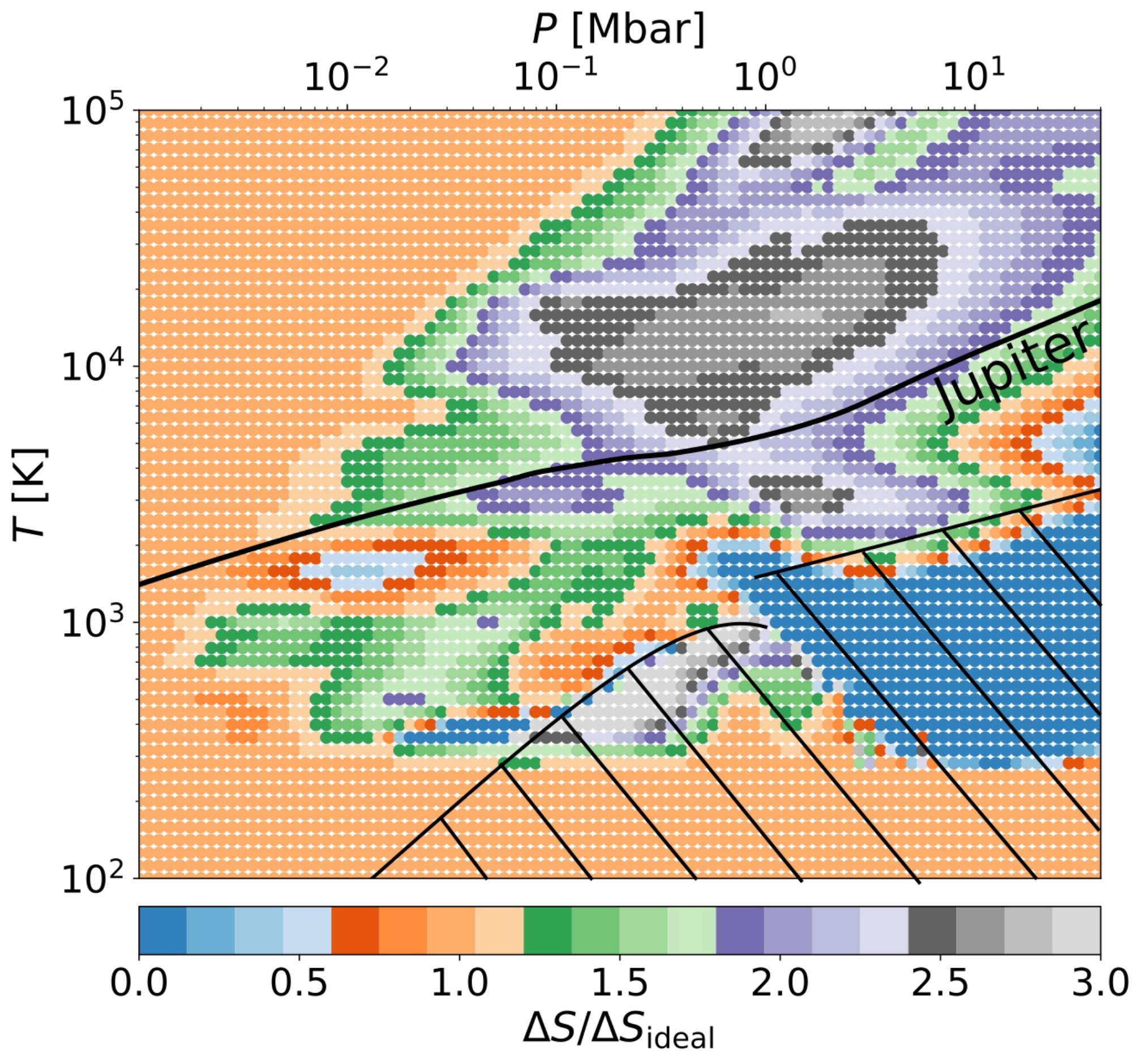}}
      \caption{Ratio of the mixing entropy $\Delta S$ and the ideal entropy of mixing $\Delta S_{\rm ideal}$ for a mixture of $Y=0.245$.
      The hashed area corresponds to the region in which the entropy of mixing is invalid \citep[see][for more details]{2019ApJ...872...51C}.
      }
         \label{figure:ds_smixan}
\end{figure}

Figure~\ref{figure:ds_smixan} shows that in the classical regime, in which $T\gtrsim 10^6 {\,\rm K}\ (P/{\rm Mbar})$, the entropy excess is equal to the ideal value from Eq.~\eqref{eq:ds_analytic}. At lower temperatures, the mixing entropy term increases to up to almost three times the ideal value, except in a small region centred on $10^{-2}\,$Mbar and $2000$\,K, where we see a decrease to only about 0.5 times the ideal value. Again, this decrease may be due to interpolation issues inherent to the original CMS19 and CD21 tables in a region that combines output from different works. For a typical Jupiter adiabat, $\Delta S$ is equal to the classical value up to $5$\,kbar and then increases to between $1.5$ and $2.5$ times this value, with a maximum around 0.5\,Mbar. It is important to realise that uncertainties on both the pure and mixed EOSs still exist, as shown by a comparison to the SR18 calculations in Appendix~\ref{section:app1}.

%-------------------------------------------------------------------
%-------------------------------------------------------------------
\section{Application to Jupiter and Saturn}
  \label{section:jupiter}
%-------------------------------------------------------------------
%-------------------------------------------------------------------

\begin{figure}
   \centering
   \includegraphics[width=\hsize]{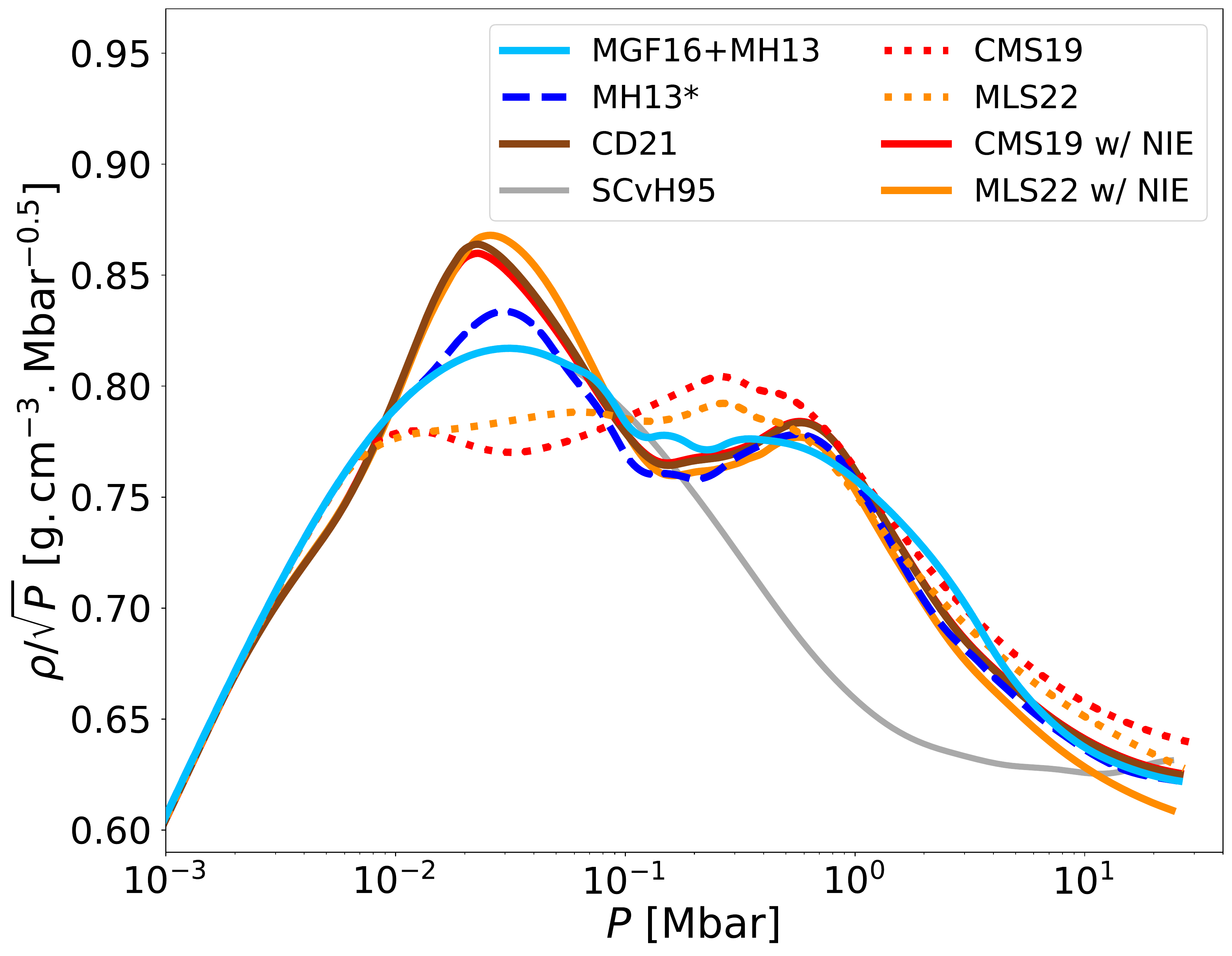}
      \caption{Adiabats obtained from different EOSs and corresponding to a homogeneous model, without a compact core, with $Y=0.245$. NIE stands for non-ideal effects.
              }
         \label{figure:eos_diversity}
\end{figure}

After deriving a table containing the mixing terms, we applied it to the cases of Jupiter and Saturn. We first compared adiabats for pure hydrogen-helium mixtures just by integrating the adiabatic gradient starting from 1\,bar, 166.1\,K (i.e. adapted to Jupiter conditions). We used a simple homogeneous model, without a compact core and with a uniform helium composition ($Y=0.245$). For convenience, because in the Jupiter interior $P \propto \rho^2$ \citep{1975SvA....18..621H}, Fig.~\ref{figure:eos_diversity} compares values of $\rho/\sqrt{P}$ obtained for different EOSs.
We note that all the new EOSs are much denser than SCvH95 \citep{1995ApJS...99..713S} in a wide region, from pressures of about 0.1 to 10\,Mbar. The CMS19 \citep{2019ApJ...872...51C} and MLS22 \citep{2022A&A...664A.112M} EOSs do not include the interactions between hydrogen and helium at all. They were obtained using Eqs.~\eqref{eq:avl_rho_dv} and \eqref{eq:avl_s_ds} with the mixing terms $\Delta V$ and $\Delta S$ both equal to 0 and combining a pure hydrogen table (CMS19-H or MLS22-H) with a pure helium table (CMS19-He or SCvH95-He). They are particularly less dense between 0.01 and 0.1~Mbar but denser at depth than other EOSs. The MGF16+MH13 \citep{2016A&A...596A.114M} and CD21 \citep{2021ApJ...917....4C} EOSs include the non-ideal mixing effects from \citet{2013ApJ...774..148M}, who evaluated the H-He interactions for a mixture with $\Ytilde=0.245$ and thus are very close to MH13* \citep{2013ApJ...774..148M}. They are also based on Eqs.~\eqref{eq:avl_rho_dv} and \eqref{eq:avl_s_ds} with $\Delta V$ and $\Delta S$ both equal to 0 ; but the non-ideal mixing effects evaluated at $\Ytilde=0.245$ are included in the pure effective hydrogen table (MGF16+MH13-H or CD21-H; see \citet{2021ApJ...917....4C} for more explanations on how to derive an effective pure H table that accounts for the H-He interactions). While MGF16+MH13-H is combined with SCvH95-He, CD21-H is combined with CMS19-He. Then, based on the table derived in Sect.~\ref{section:derivation}, we added the mixing terms to the CMS19 and MLS22 EOSs and obtained adiabats (CMS19 w/ NIE and MLS22 w/ NIE) that are closer to the other EOSs that include the non-ideal mixing effects (MGF16+MH13, MH13*, and CD21).

\begin{figure}
   \centering
   \includegraphics[width=\hsize]{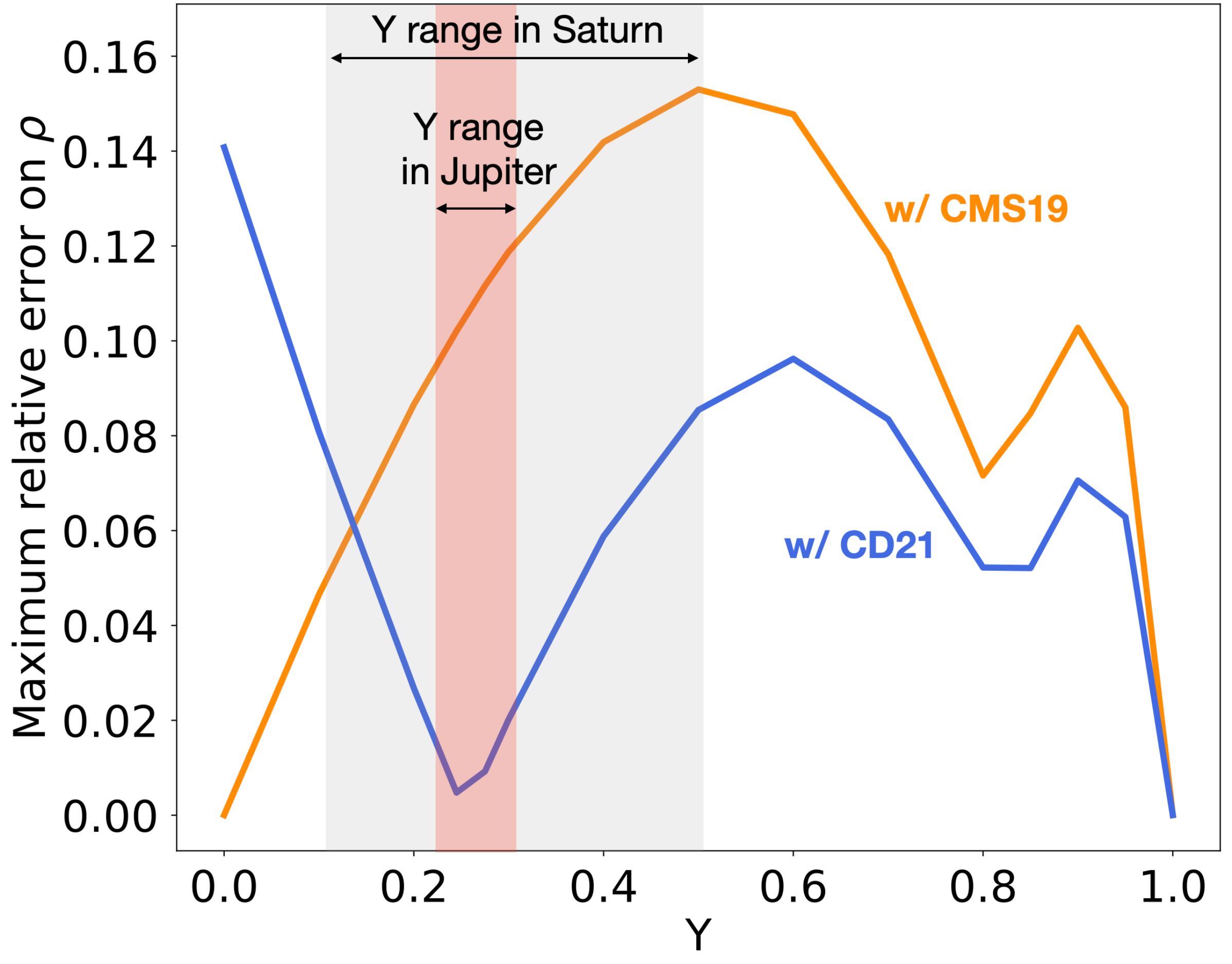}
      \caption{Comparison in density of our new EOS (this work) based on the non-ideal mixing effects derived in Section~\ref{section:derivation} with both CD21 and CMS19, according to the helium mass fraction $Y$. The shaded red and grey areas correspond to the range of values of $Y$ in Jupiter (taken from models of \citet{2022A&A...662A..18M}) and Saturn (taken from \citet{2021NatAs...5.1103M}), respectively.
              }
         \label{figure:max_dev_Y}
\end{figure}

An EOS that does not properly include the interactions between hydrogen and helium tends to underestimate the total amount of heavy elements in Jupiter. A quick test that consists of estimating the amount of heavy elements needed to match the equatorial radius of Jupiter, using a static model, can give a rough estimate of the impact of taking the non-ideal mixing effects into account. Focusing on a model similar to the one used for the adiabats of Fig.~\ref{figure:eos_diversity}, with $Y=0.245$, we ran two models: one model with CMS19, and one model with CMS19, including the non-ideal mixing effects. In the first case, we needed to add a compact core of 14.1~$M_{\oplus}$ to fit the equatorial radius of Jupiter, while we needed to add a compact core of 20.8~$M_{\oplus}$ in the second case. This shows that using the CMS19 EOS \citep[see][]{2019A&A...632A..76N,2021PSJ.....2..241N,2022A&A...662A..18M} leads to an underestimation of the amount of heavy elements in Jupiter.

Another issue arises when the helium mixing ratio departs from the value $\Ytilde=0.245$, and particularly, as in Jupiter and Saturn, when the helium abundance varies from the protosolar value due to H-He phase separation \citep{1977ApJS...35..239S}. In this case, even EOSs based on \citet{2013ApJ...774..148M} such as MGF16+MH13 and CD21 continue to evaluate the non-ideal mixing effects at $\Ytilde=0.245$. In order to evaluate the magnitude of this effect, we proceeded as follows. First, we calculated some adiabats (at 1\,bar conditions for Jupiter) for several values of $Y$ and for the three EOSs CMS19, CD21, and our new EOS (CMS19 w/ NIE), which is based on the combination of CMS19-H, CMS19-He, and the mixing terms derived in Sect.~\ref{section:derivation}, using Eqs.~\eqref{eq:avl_rho_dv} and \eqref{eq:avl_s_ds}. We then searched for the maximum difference in density between the adiabats derived for each EOS at every value of $Y$. Figure~\ref{figure:max_dev_Y} shows that the maximum deviation between our new EOS and CMS19 reaches a peak near $Y=0.5$ and is equal to 0 at $Y=0$ and $Y=1$ because our new EOS is constructed to recover the pure end members CMS19-H and CMS19-He. With CD21, the maximum deviation reaches a minimum near $Y=0.245$, which corresponds to the value where the non-ideal mixing effects were estimated. (The minimum is not exactly zero because of the difference between $\Ytilde=0.245$ and $\Ystar=0.275$.) In the case of Jupiter, a relative difference of 12\% can exist (for $Y$ close to 0.3) between our new EOS and CMS19. The comparison with CD21 yields a 2.5\% difference in density at most. For Saturn, where the range of $Y$ is wider \citep{2021NatAs...5.1103M}, the effects of the H-He interactions are even stronger. Using CMS19 can lead to a 15\% error in density (for $Y=0.5$), while using CD21 can lead to an error of 8.5\%. 

%-------------------------------------------------------------------
%-------------------------------------------------------------------
\section{Application to exoplanets}
  \label{section:exoplanets}
%-------------------------------------------------------------------
%-------------------------------------------------------------------

\begin{figure}
   \centering
   \includegraphics[width=\hsize]{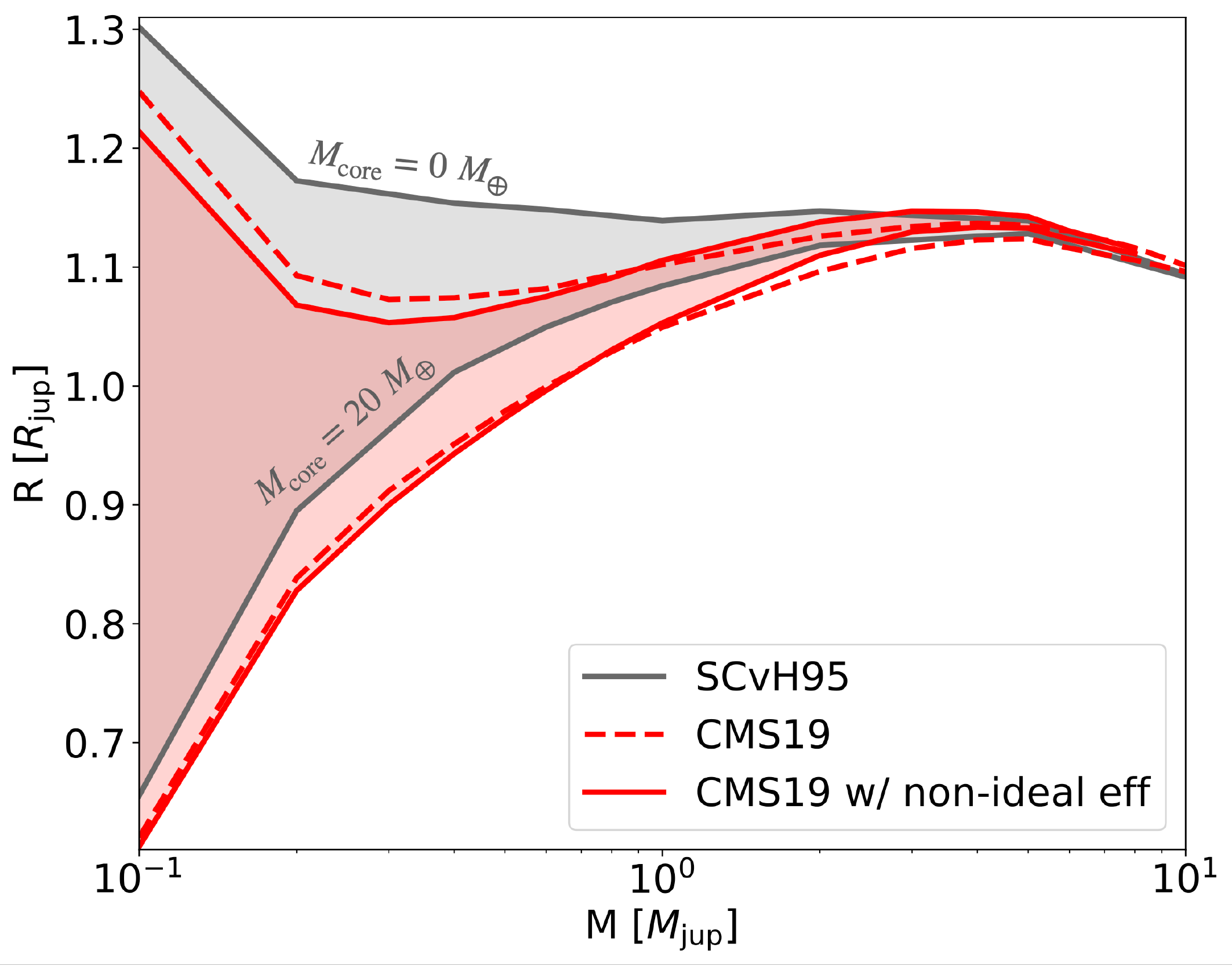}
      \caption{Evolution models of giant planets for three EOSs: SCvH95, CMS19, and our new EOS (CMS19 w/ non-ideal effect), with $T_{\rm eq}$=1500~K, taken at 4.5~Gyr. The models consist of a pure rocky core overlaid with a H-He envelope of solar composition. The coloured area for each EOS is bounded by evolution models without a core and with a 20~$M_{\oplus}$ core. 
              }
         \label{figure:exoplanets}
\end{figure}

So far, we focused on static models to analyse internal structures at a given time. Now, we use CEPAM and a non-grey atmosphere \citep{2006A&A...453L..21G,2015A&A...574A..35P} to model the evolution of planets and determine the influence of the H-He interactions during their lifetime. As seen in Sect.~\ref{section:jupiter}, when we do not include non-ideal mixing effects, static models lead to an underestimated amount of heavy elements, thus a denser mixture of hydrogen and helium overall. Hence, static models (with similar quantities of heavy elements) will yield a larger radius when accounting for these non-ideal mixing effects. Assuming simple structures consisting of a central rocky core and a surrounding H-He envelope of solar composition ($Y=0.3$), we calculated evolution models for giant planets with $T_{\rm eq}$=1500~K. For each EOS, Figure~\ref{figure:exoplanets} shows the mass-radius relation obtained at 4.5~Gyr, bounded by homogeneous models without a core (on top) and with a core of 20~$M_{\oplus}$ (at the bottom). Static models (for fixed boundary conditions) yield larger radii when non-ideal mixing effects are included, by up to 2\% for planets of 1-3~$M_{\rm Jup}$. When we consider the evolution, the present-day radius results both from the mass-radius relation for static models and from the history of the planet in the $P-T$ diagram. Figure~\ref{figure:exoplanets} shows that the effects are generally in the opposite sense, leading to a change in radius due to non-ideal mixing that is smaller than for purely static models. We find larger radii for planets that are more massive than Jupiter when we use our new EOS (CMS19 w/ NIE) compared to CMS19; and in contrast to static models, we find smaller radii (by up to 2\%) for planets that are less massive than Jupiter. The difference can reach 6\% at most, at younger ages. Furthermore, we stress that the radii differ significantly (by up to 8\%) from those obtained with the SCvH95 EOS, which is commonly used \citep{2007ApJ...659.1661F}. Because of the accuracy in masses and radii of exoplanets, more careful modelling is needed and improved knowledge of the H-He EOS is important to characterise exoplanets.

%-------------------------------------------------------------------
%-------------------------------------------------------------------
\section{Conclusion}
%-------------------------------------------------------------------
%-------------------------------------------------------------------

We provide a simple way to account for the interactions between hydrogen and helium in the EOS. We have derived mixing terms (including the non-ideal contribution) to be added to the classically used linear mixing law for density and entropy, respectively. We built a table of these mixing terms as a function of pressure and temperature. This table is to be combined with pure EOSs to derive the properties of any hydrogen and helium mixture. We emphasise that omitting these mixing effects can lead to a relative error in density of up to 15\% on the H-He EOS for Jupiter and Saturn, resulting in a serious underestimation of the heavy element reservoir and invalidating the fit to the observed gravitational moments. Even models including these non-ideal mixing effects evaluated for a helium mass-mixing ratio $Y=0.245$ \citep{2019ApJ...872..100D,2022A&A...662A..18M,2022PSJ.....3..185M,2021NatAs...5.1103M} have intrinsic errors on the density evaluated to up to 2.5\% for Jupiter and 8.5\% for Saturn. In addition, when the non-ideal mixing effects are not included, a change of up to 6\%  in the calculated radii of exoplanets may result. Furthermore, given the differences between EOSs, we stress that uncertainties on the H-He EOS still exist. Our approach provides a way to obtain a self-consistent EOS for different helium abundances. We stress that it relies on EOSs calculated for three compositions only (for $Y=0$, $Y=0.245$ and $Y=1$). Ab initio calculations and high-pressure experiments for $P=10^{-2}-10$\,Mbar and $T=1000-20,000$\,K with other hydrogen-helium compositions, but also including heavy elements, would be highly desirable.

\begin{acknowledgements}
      We warmly thank R. Redmer and A. Bergermann for sharing their EOS data. We also thank the referee for constructive comments which improved the quality of the manuscript. We thank the Juno Interior Working Group and particularly R. Helled, S. Müller and N. Nettelmann for useful discussions and comments. This research was carried out at the Observatoire de la Côte d’Azur under the sponsorship of the Centre National d’Etudes Spatiales.
\end{acknowledgements}

   \bibliographystyle{aa} % style aa.bst
   \bibliography{aanda} % your references Yourfile.bib

\begin{appendix} %First appendix

\section{Comparisons of the entropy of mixing}
    \label{section:app1}

\begin{figure}
   \centering
   \includegraphics[width=\hsize]{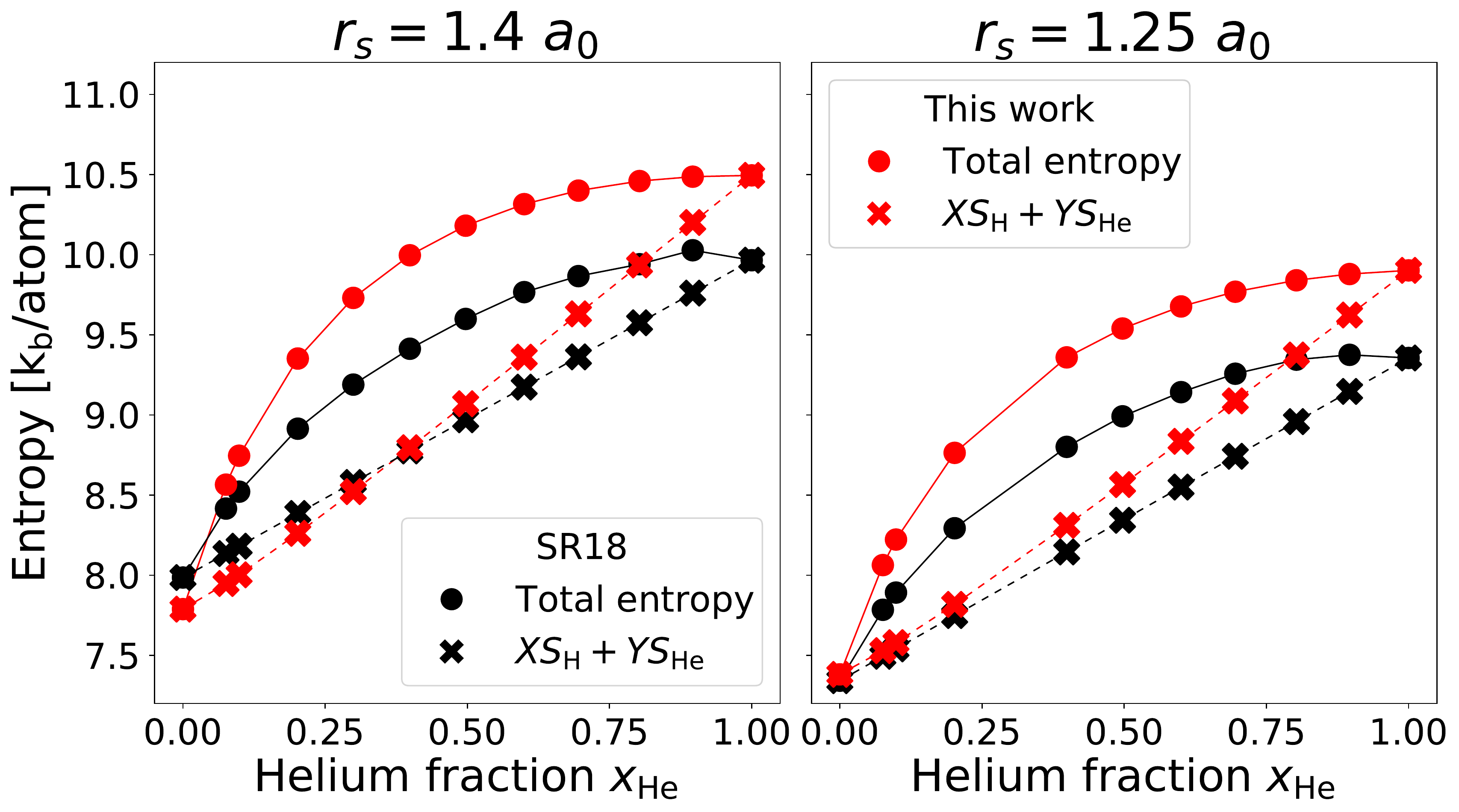}
      \caption{Comparison of our entropy with the one calculated by SR18 (from Fig.1 of their supplemental material) for $T=10 000~K$ and various helium fractions $x_{\rm He}$. The two panels show results for values of Wigner-Seitz radii of $r_{\rm s}=1.4 \, a_0$ and $r_{\rm s}=1.25 \, a_0$ respectively. Each panel displays our calculations (\textit{red}) of the total entropy (computed using Eq.~\ref{eq:avl_s_ds} with CMS19-H, CMS19-He and our table for $\Delta S$) and also the entropy without including the entropy of mixing $\Delta S$ (hence corresponding to $XS_{\rm H}+YS_{\rm He}$). Results from SR18 are shown in \textit{black}.
              }
         \label{figure:full_entropy}
\end{figure}

Figure~\ref{figure:full_entropy} shows a comparison of the entropies obtained from our work and by SR18 (SM Fig.1). For $T=10 000~K$, $r_{\rm s}=1.4 \, a_0$ and $1.25 \, a_0$, and various helium fractions $x_{\rm He}$, we calculated the total entropy (using Eq.~\eqref{eq:avl_s_ds}, with CMS19-H, CMS19-He, and our table of the mixing terms to compute $S_{\rm H}$, $S_{\rm He}$ and $S_{\rm mix}$, respectively). We also show the entropy resulting from the ideal mixing of only the pure terms ($XS_{\rm H}+YS_{\rm He}$). The difference between the two curves represents the entropy of mixing $\Delta S$. First, we note some important differences on the pure H and He EOSs between CMS19 and the work of SR18. Typically, for $r_{\rm s}=1.4 \, a_0$, the entropy of CMS19-H is $0.2~\rm k_b/atom$ lower than the pure H EOS from SR18, while the entropy of CMS19-He is $0.53~\rm k_b/atom$ higher than the pure He EOS from SR18. In H-He mixtures, the entropy we calculate (where the mixing entropy is initially based on the MH13 EOS) is always higher than the entropy from SR18 ($0.58~\rm k_b/atom$ greater at most).

These intrinsic differences in the pure EOSs and also in the entropies yielded for H-He mixtures lead to discrepancies between the entropy of mixing that we obtain and the one from SR18. In this parameter space ($T=10 000~K$, $r_{\rm s}=1.4 \, a_0$ or $1.25 \, a_0$), while the entropy of mixing of SR18 ranges from 0.23 to $0.66~\rm k_b/atom$, our entropy of mixing ranges from 0.26 to $1.2~\rm k_b/atom$. The entropy of mixing that we obtain is hence 1.1 to 2.2 times higher than the entropy of mixing of SR18. (Similar results would be obtained using the \citet{Morales2009} calculations because their entropies of mixing are similar to those of SR18). This highlights the fact that the EOSs for pure hydrogen, pure helium, and mixtures of hydrogen and helium obtained by the different groups still differ; this is particularly obvious for the pure helium EOS used here (CMS19-He) and the one obtained by SR18. While CMS19 have used the Perdew–Burke–Ernzerhof (PBE) exchange-correlation functional in their simulations, SR18 applied the van der Waals density functional (vdW-DF) exchange-correlation functional. Both exchange-correlation functionals support different experiments and have their own benefits and limitations (see SR18 for a more detailed discussion). A consistent, homogeneous exploration of EOSs in this regime of pressures and temperature and for different compositions is crucial.

In spite of these uncertainties, we believe that using the table of non-ideal mixing effects that we provide in addition to the CMS19 EOS is an improvement for two reasons: (i) It includes non-ideal mixing effects that are otherwise ignored, and for low helium abundances (Y~<0.3), it yields total entropies that differ from those of SR18 by less than 4\%. (ii) It enables calculations in the high-temperature regions spanned by massive planets and brown dwarfs, which are otherwise not possible with the CD21 EOS.

\section{Table of the non-ideal mixing effects}
    \label{section:app2}

\begin{table*}[hbt]
\caption{Table of the mixing terms.}            
\label{table:1}      
\centering          
\begin{tabular}{c c c c l l l }     % 7 columns 
\hline\hline       
                      % To combine 4 columns into a single one 
log($P$) [dyn/cm$^2$] & log($T$) [K] & $V_{\rm mix}$ [cm$^3$/g] & $S_{\rm mix}$ [erg/g/K] \\ 
\hline                    
10.5 & 2.5 & 0.21972032499215227 & 145270983.02864826 \\
10.5 & 2.55 & -1.6752656793119542 & 97266054.24784786 \\
10.5 & 2.6 & -0.5750688394019108 & 19976893.840987258 \\
10.5 & 2.65 & -0.20048821618796142 & 114994434.5574714 \\
10.5 & 2.7 & 0.013568193301054745 & 157590130.69068393 \\
10.5 & 2.75 & 0.1328211678295867 & 153264410.38666478 \\
10.5 & 2.8 & 0.21204647603740195 & 153310129.47913098 \\
10.5 & 2.85 & 0.24501465036396686 & 156237499.56270728 \\
\hline
\end{tabular}
\tablefoot{A full version of the electronic table is available via \url{https://doi.org/10.5281/zenodo.7346181}.
}
\end{table*}

The derivation of $V_{\rm mix}$ and $S_{\rm mix}$ is described in Sect.~\ref{section:derivation}. The table had to be slightly adjusted in order to avoid spurious interpolation issues and to be directly usable for interior and evolution models of giant planets and brown dwarfs. $V_{\rm mix}$ displayed an alternation of positive and negative values with very low amplitude that jeopardised the interpolation through the table. Hence $V_{\rm mix}$ was set to 0 when $\Delta V \times \rho_{\rm H-He}$ was lower than 0.01\%. Thirteen isolated outliers, located in a region of the table where 11.1 < log($P$) < 12.5 (with pressure in dyn/cm$^2$) and 2.45 < log($T$) < 2.85 (with temperature in K), were smoothed with a median over the contiguous points of the table. $S_{\rm mix}$ presented strong outliers at the edges of the table both at low (log($P$) < 6) and high pressure (log($P$) > 15), with values that could differ by several orders of magnitudes. Hence, for regions of the table where 6 < log($P$) < 15, we set the values of $S_{\rm mix}$ so that $\Delta S$ is equal to the ideal entropy of mixing. Furthermore, when we used the table to model the evolution of massive brown dwarfs ($\sim 60~M_{\rm J}$) using our planetary evolution code, we ran into convergence issues. We found that the CD21 entropy table caused this at high temperatures (log($T$) > 5). We therefore multiplied $\Delta S(\Ytilde) - \Delta S_{\rm ideal}(\Ytilde)$ by $\left\{1 - \rm{erf} \left[(T-T_{\rm ref})/\delta T\right] \right\}/2$ with $T_{\rm ref}=4.9$ and $\delta T=0.1$: At higher temperatures, the entropy of mixing smoothly converges to the ideal value.

A fragment of the final version is presented in Table~\ref{table:1}. The table may be used both in the low- (giant planets) and in the high-mass regime (brown dwarfs). We recall that this table is valid in the same domain as the EOS from \citet{2019ApJ...872...51C}, whose limitations concern the regions in which molecular hydrogen and ionised hydrogen become solid, as well as in the region in which ion quantum effects become important, namely at low T and high P (see Fig.1 and Fig.16 of \citet{2019ApJ...872...51C} for the precise locations in the phase diagrams).

\end{appendix}
%---------

\end{document}